\documentclass[12pt]{article}

\usepackage{sbc-template}

\usepackage{graphicx,url}

\usepackage[utf8]{inputenc}  
\usepackage{multirow}
     
\title{What Developers Ask to ChatGPT in GitHub Pull Requests? an Exploratory Study}

\author{Julyanara R. Silva\inst{1}, Carlos Eduardo C. Dantas\inst{1}, Marcelo A. Maia\inst{2}}

\address{Instituto Federal de Ciência e Tecnologia do Triângulo Mineiro (IFTM) \\ Campus Uberlândia Centro -- Uberlândia, MG -- Brazil
\nextinstitute
  Universidade Federal de Uberlândia (UFU) -- Uberlândia, MG -- Brazil
\email{\{julyanara.silva@estudante.,carloseduardodantas@\}iftm.edu.br, marcelo.maia@ufu.br}
}

\begin{document} 

\maketitle

\begin{abstract}

The emergence of Large Language Models (LLMs), such as ChatGPT, has introduced a new set of tools to support software developers in solving programming tasks. However, our understanding of the interactions (i.e., prompts) between developers and ChatGPT that result in contributions to the codebase remains limited. To explore this limitation, we conducted a manual evaluation of 155 valid ChatGPT share links extracted from 139 merged Pull Requests (PRs), revealing the interactions between developers and reviewers with ChatGPT that led to merges into the main codebase. Our results produced a catalog of 14 types of ChatGPT requests categorized into four main groups. We found a significant number of requests involving code review and the implementation of code snippets based on specific tasks. Developers also sought to clarify doubts by requesting technical explanations or by asking for text refinements for their web pages. Furthermore, we verified that prompts involving code generation generally required more interactions to produce the desired answer compared to prompts requesting text review or technical information.

\end{abstract}

\section{Introduction}

In recent years, several large-scale language models (LLMs) based on the Transformer architecture have emerged \cite{vaswani2017attention}. A notably recent and prominent model is ChatGPT \footnote{https://openai.com/blog/chatgpt}, which is a state-of-the-art language model developed by OpenAI \cite{Radford2019}. Although ChatGPT was designed for natural language processing tasks, it also offers potential in providing solutions in software development environment. For example, developers can ask ChatGPT for code to solve a specific programming task or provide a detailed explanation of some concept \cite{Ernst_2022}. 

Despite the relatively recent emergence of LLMs, their growing use in repositories such as GitHub have been documented. There are numerous mentions of ChatGPT in commits, Pull Requests (PRs), and even source code comments where ChatGPT provided a solution that was employed by developers \cite{devgpt}. Recent studies have also demonstrated that ChatGPT has been particularly useful for performing programming tasks such as refactoring operations into their source code, fix bugs, and others \cite{tufano2024}, presenting an alternative by generating possibly original answers, including modified versions from developers source code \cite{Ebert2023}.

However, some previous works evaluated the use of ChatGPT in a more general manner, i.e., in commits, issues, (open, closed and merged) PRs, discussions, which results in a mixture of ChatGPT usage that do not specifically focus on contributions to the codebase. For instance, by evaluating the use of ChatGPT specifically for merged PRs in a peer code review process, we are focusing on scenarios where the proposed changes by developers were generally accepted by the reviewers. This approach ensures that we are capturing instances where ChatGPT assistance could led to tangible contributions in the GitHub repository, thereby providing a more specific and refined sample. 

Another limitation to address involves in effectively evaluating the interactions (i.e., prompts) between developers and the ChatGPT to generate the desired solution, as single-query responses from LLMs might not fully address the developer's needs. LLMs mantains a context window, and these interactions involve refining and expanding the dialogue until the model generates an accurate and helpful response \cite{Ebert2023,Mondal2024}. We envision that understanding what developers ask in these prompts and evaluate the number of prompts may help leveraging the use of LLMs in programming tasks.

In this study, our objective is to evaluate the prompts performed by developers and reviewers in 155 valid ChatGPT share links extracted from 139 merged PRs across 115 GitHub repositories, revealing the interactions between developers and reviewers with ChatGPT. Our study provides the following contributions:

\begin{enumerate}
    \item A catalog of 14 types of requests performed on ChatGPT, categorized into four main groups. This catalog provides insights into the various types of interactions between developers and ChatGPT, particularly focusing on their collaboration to find solutions to programming tasks.
    \item An analysis of how the number of prompts required to find the solution varies across different categories of requests made by developers and reviewers to ChatGPT.  
    \item The replication package that includes the data and scripts, which can be used for future research  \cite{carlos_eduardo_c_dantas_2023}. 
\end{enumerate}

The paper is organized as follows. Section 2 presents the methodology to answer the research questions. The results are reported and discussed in Section 3. Section 4 presents the threats that could affect the validity of this study. Section 5 presents the related literature. Finally, Section 6 summarizes our observations and outlines directions for future work.

\section{Methodology}

The study is driven by two main research questions:

\begin{itemize}

\item \textbf{RQ \#1) What do developers request on ChatGPT to solve Pull Requests?} This RQ aims to understand and categorize the prompts provided by developers to ChatGPT, using the share links cited in the merged PRs. Therefore, instead of searching for mentions of ``ChatGPT" keyword in GitHub as done in previous works \cite{tufano2024}, we focused only on merged PRs where developers shared links exposing their interactions with ChatGPT.

\item \textbf{RQ \#2) What is the distribution of the size of the interaction (\#prompts) with ChatGPT to get the desired answer?} This RQ aims to examine the iterative process between the developer and ChatGPT, focusing on the dialogue until the model generates an accurate and helpful response. The quality of the prompts influences developers' performance during programming tasks \cite{Ebert2023,Mondal2024}.
\end{itemize}

In the following subsections, we detail how we selected the samples (Section 2.1), performed the manual evaluation (Section 2.2), and obtained the distribution of the prompts (Section 2.3). 

\subsection{Mining Candidate Merged Pull Requests (PRs)}

The goal of this step is to identify potential merged PRs where ChatGPT was likely used to assist developers. We began by writing a script that utilized the GitHub GraphQL API\footnote{https://docs.github.com/en/graphql} to query for non-forked merged PRs, mentioning the ChatGPT share link: ``chat.openai.com/share". We first executed this query in May 2024 to collect and work with the initial samples and then executed it again in July 2024 to capture any additional samples, completing the dataset. To avoid toy repositories, we filtered for non-forked repositories with at least 10 stars, as defined in previous work \cite{Dabic2021SamplingPI}. The output of this step returned 302 candidate merged PRs.

Next, we performed manual filtering to discard as many false positives as possible: 

\begin{enumerate}
\item 97 merged PRs where the ChatGPT share link was not found in the PR title, body, comments, commit messages or code diffs.
\item 29 merged PRs with broken or invalid ChatGPT share links (e.g., error 404).
\item 25 merged PRs written in non-english languages.
\item 12 merged PRs without any reviewer (e.g., the developer themselves opened the PR and merged it without receiving any feedback on their modification)

\end{enumerate}
 
\begin{figure}[ht]
\centering
\includegraphics[width=1.0\textwidth]{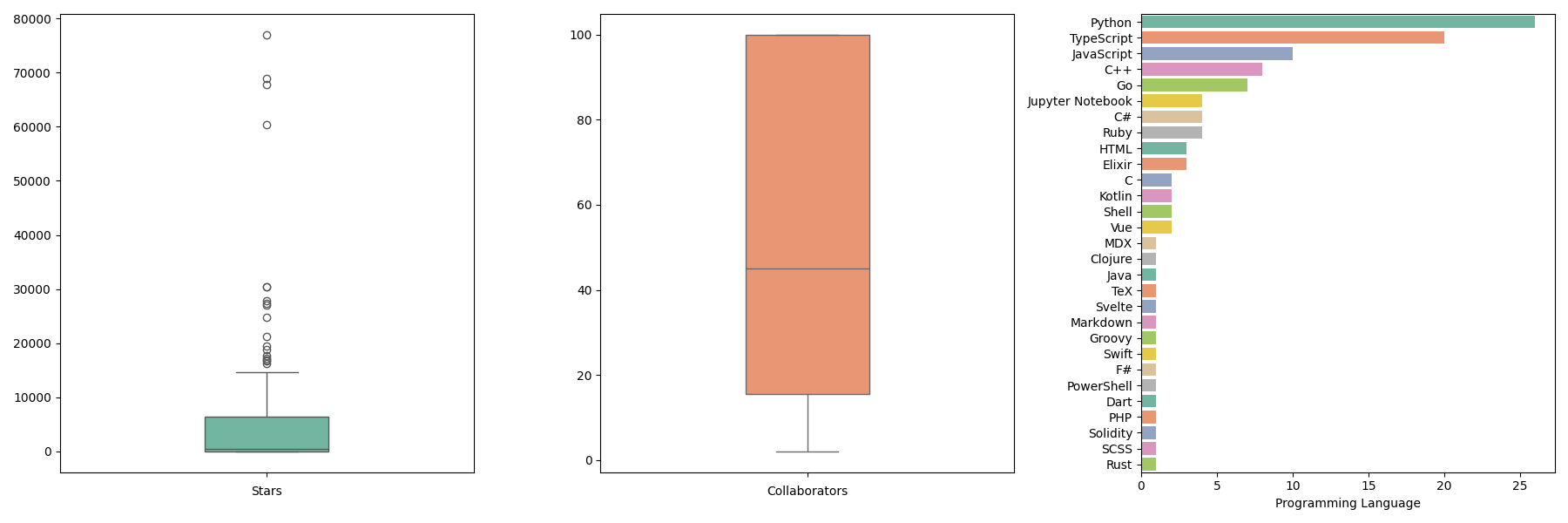}
\caption{Distribution of number of stars, contributors and programming languages among the selected repositories.}
\vspace{-4mm}
\label{fig:boxplot}
\end{figure}

This process resulted in a final set of 155 valid ChatGPT share links (instances) from 139 merged PRs across 115 distinct repositories. Figure \ref{fig:boxplot} illustrates the distribution of stars, contributors and main programming languages among these repositories. Although it might be expected to find more candidate PRs, our work required specific filters, as mentioned. For comparison, the DevGPT dataset, in its latest snapshot (2023-10-12), includes 265 PRs, which do not include the specific filters applied in our work \cite{devgpt}.

\subsection{Characterizing the Requests for ChatGPT Assistance}

This step aims to address RQ \textit{\#1} by characterizing what developers requests for ChatGPT assistance in the 155 instances found in merged PRs. Two evaluators (authors of this paper) conducted the manual inspection. Each instance was independently reviewed by each evaluator, analysing the links: (i) a link to the GitHub merged PR (ii) a link to the ChatGPT share link provided inside the PR. 

\begin{figure}[ht]
\centering
\includegraphics[width=0.6\textwidth]{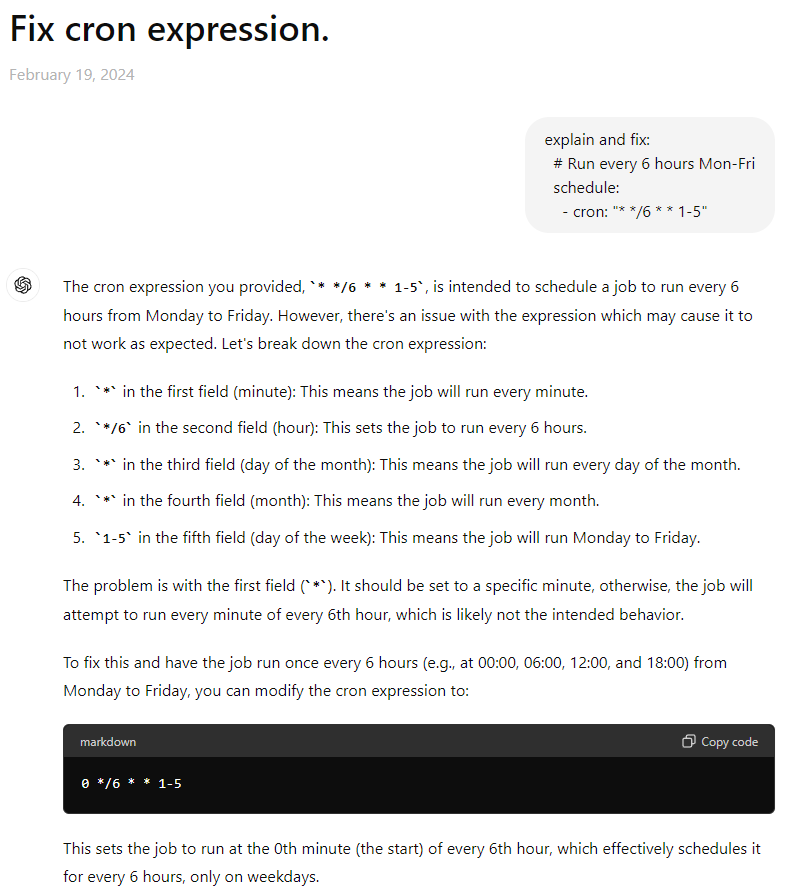}
\vspace{-5mm}
\caption{Example of ChatGPT prompt extracted from Merged PRs}
\vspace{-4mm}
\label{fig:example}
\end{figure}

The categorization involve separately assigning one or more labels to classify the types of requests using the open coding process \cite{spencer2009card}. These labels represent the main topics developers require in the ChatGPT prompts. Figure \ref{fig:example} illustrates an example. In this case, the extracted labels were \textit{explain the source code} and \textit{fix the source code}, as ChatGPT provided an explanation of what the code is doing and fixed the error by setting the job to run once every six hours from Monday to Friday. These labels result in two distinct candidate types of requests.  To maintain consistency in labeling, evaluators shared and reviewed their labels in a common repository throughout the analysis.

After the first round, the evaluators met to revise the set of defined labels, merging similar labels and renaming unclear ones. For example, one evaluator created two distinct labels for fixing warnings and fixing errors, and shared these with another evaluator. The other evaluator preferred to merge fixing warnings and errors into a single label. After the meeting, both evaluators agreed that fixing warnings and errors were similar approaches and could be merged into one label. 

The evaluators also found conflicts in open coding category for 25 cases (approximately 15\% of instances). This percentage was expected because we evaluate a large set of prompts across many different programming languages and technologies, leading to discrepancies in extract the labels. Some related works generally produces qualitative analyses in specific topics such as readability \cite{Dantasicsme}, refactoring \cite{Silva2016,Pantiuchina2020} and others. These cases were reclassified into new topics during the revision process. For example, some instances previously labeled as Technical Explanation were re-evaluated by both evaluators and agreed to be better categorized as Technical Support, resulting in a new type of ChatGPT request.

The 14 labels defined through the above-described process were used to categorize the types of ChatGPT requests made by developers. The evaluators then grouped similar types of requests with analogous purposes to create the initial taxonomy. This first draft described the various purposes of ChatGPT requests, such as \textit{code review}, \textit{text review}, etc. The authors subsequently refined the taxonomy by renaming certain categories and reorganizing sub-categories to enhance clarity. The final version consists in four main categories.

\subsection{Distribution of the prompts with ChatGPT}

%To address RQ \textit{\#2}, we extracted the number of prompts provided by developers in each category (Code Review, Information Request, Code Generation and Text Review) and compared the distribution of the dependent variable (number of prompts) across the independent variable (ChatGPT request category). We applied analysis of variance (\textit{ANOVA}) with a 5\% confident level (i.e., \textit{p-value $<$ 0.05}) to determine if there were any statistically significant differences. To identify which specific categories were significantly different from each other, we conducted a post-hoc analysis using the \textit{Tukey test}.

To address RQ \textit{\#2}, we extracted the number of prompts provided by developers in each category (Code Review, Information Request, Code Generation, and Text Review) and compared the distribution of the dependent variable (number of prompts) across the independent variable (ChatGPT request category). We applied the Kruskal-Wallis test with a 5\% confidence level (i.e., \textit{p-value $<$ 0.05}) to determine if there were any statistically significant differences, as the Shapiro-Wilk test indicated non-normality in the distributions. To identify which specific categories were significantly different from each other, we conducted a post-hoc analysis using the \textit{Dunn test}.

\section{Results}

The results are driven to answer each research question previously mentioned.

%\subsection{RQ \#1) Why developers ask ChatGPT assistance in the context of GitHub Pull Requests?}

\subsection{RQ \#1) What do developers request on ChatGPT to solve Pull Requests?}

Table \ref{tab:rq1} presents 14 types of requests found in the analyzed ChatGPT responses extracted from merged PRs, categorized into four distinct categories. It is important to note that the total count of instances across all tasks (163) exceeds the total number of valid instances inspected (155). This discrepancy arises because a single instance may have required ChatGPT's support for multiple tasks, as illustrated in Figure \ref{fig:example}, such as both explaining and fixing the source code.

\begin{table*}[htbp]
\centering
\caption{ChatGPT requests performed in Merged Pull Requests}
\label{tab:rq1}
\begin{tabular}{l|l|c|c}
\hline
Category & ChatGPT Request & Occurrences & Total \\
\hline
\multirow{ 3}{*}{Code Generation} 
               & How-To Code Snippets & 30 & \multirow{ 3}{*}{61}  \\  
               & Task Automation Requests & 20 \\
               & Feature Addition to Existing Code & 11  \\
\hline                         
\multirow{ 5}{*}{Code Review}                
               & Fix Bugs and Warnings  & 18 & \multirow{ 5}{*}{46}  \\   
               & Optimization/Refactoring & 13 \\
               & Explain the Code & 8 \\                   
               & Test/Debug & 5 \\
               & Performance Analysis & 2 \\
\hline   
\multirow{ 4}{*}{Information Request} 
               & Technical Explanation & 24 & \multirow{ 4}{*}{42}  \\    
               & Technical Support & 10 \\                              
               & Coding Conventions & 5 \\             
               & Policy & 3 \\                
\hline  
\multirow{ 2}{*}{Text Review} 
               & Grammar and Refinement & 10 & \multirow{ 2}{*}{14}  \\                 
               & Formatting & 4  \\
\hline                   
\multicolumn{3}{c}{\textbf{Total}} & \textbf{163}  \\
               
\hline

\end{tabular}
\end{table*}

\subsubsection{Code Generation} 

This category involves asking ChatGPT to generate code based on the requirements provided in the prompt. The requirement can be a textual description of what needs to be done (e.g., ``how to do") or a code snippet requesting ChatGPT to implement something similar or create a new functionality based on that snippet. 

\textit{How-To Code Snippets} (30 out of 61 samples) - In this type of request, developers asks ChatGPT to produce a code snippet to accomplish a specific task, generally using ``how to do" questions. They may also provide JSON, XML, or database schemas, requesting code to manipulate this data. For example, \textit{PR \#4337} from \textit{temporalio/temporal} repository, ChatGPT was asked to create a script that searches for camel case file names and renames them to snake case. The changed files were then committed into the merged PR.

\textit{Task Automation Requests} (20 out of 61 samples) -  Developers input source code into ChatGPT to automate changes and avoid manual effort.  For example, in \textit{PR \#50} from \textit{pollen-robotics/rustypot}, a link to a manual containing the control table RAM of a device was provided to ChatGPT, along with a script template. ChatGPT replaced the values with the correct RAM addresses. In other instances, developers input HTML text, requesting it to be translated, or ask ChatGPT to create test cases based on their code.

\textit{Feature Addition to Existing Code} (11 out of 61 samples) - Developers provide a code snippet and ask ChatGPT to implement a new feature based on it. Typically, they request ChatGPT to rewrite their code to incorporate the new feature. For example, \textit{PR \#5937} from \textit{detekt/detekt} repository provided ChatGPT with a Kotlin class that checks for complex conditional expressions.

\subsubsection{Code Review}

In this category, developers input source code into ChatGPT to act as a code reviewer, analyzing and improving aspects of the code. Typically, the requests involve optimizing performance, identifying bugs, enhancing readability, or validating specific functionalities. Developers seek either to confirm the correctness of their code or to request modifications to improve it.

\textit{Fix Bugs and Warnings} (18 out of 46 samples) - Developers input source code into ChatGPT to fix issues, warnings, syntax errors, or unexpected crashes.  For example, in \textit{PR \#50} from \textit{poki/netlib}, the developer provided a database script asking to fix an error. ChatGPT identified the bug and rewrote the script to correct it. Another example is a cron expression fixed, as illustrated in Figure  \ref{fig:example}.

\textit{Optimization/Refactoring} (13 out of 46 samples) - Developers input source code into ChatGPT to optimize performance and perform refactoring operations such as improve clarity, naming and clean up. In \textit{PR \#16244} from \textit{betagouv/beta.gouv.fr}, the developer provided a nginx configuration asking to optimize the code. ChatGPT removed unnecessary redirections, improve performance and clarity by rewriting directives with return statements, and improve the client-side caching.

\textit{Explain the Code} (8 out of 46 samples) - Developers ask ChatGPT to explain each command in their source code.  In \textit{PR \#177} from \textit{roboflow/supervision} repository, a reviewer used ChatGPT to compare the code implemented by the developer who opened the PR with a previous version, to check for significant differences in the results. ChatGPT's response was used to suggest changes in the PR.

\textit{Test/Debug} (5 out of 46 samples) - Developers ask ChatGPT to test or debug their code by providing input values and generating output. For example, in \textit{PR \#2} from \textit{chitalian/gptask} repository, the developer submitted a Python script but noted in the PR description that it was not tested yet. The reviewer then asked ChatGPT to produce test commands for the script and shared the response in the PR.

\textit{Performance Analysis} (2 out of 46 samples) - Developers request ChatGPT to evaluate the performance of their code, focusing on time or space complexity. For example, in \textit{PR \#5068} from \textit{darklang/dark}, the developer provided the code snippet, and the ChatGPT evaluate the performance in time and complexity.

\subsubsection{Information Request} 

In this category, developers generally ask ChatGPT to explain content, much like a teacher. This content can range from explanations of computing concepts, security policies, technical support, and code conventions for specific programming languages.

\textit{Technical Explanation} (24 out of 42 samples) - Developers ask technical questions related to the software development environment, seeking answers from ChatGPT. These responses are typically used as arguments in PR discussions between developers and reviewers or to justify modifications within the PR. Instead of ``how to" questions asking for code examples, developers commonly ask ``what is" questions to better understand concepts.  For example, in \textit{PR \#30321} from \textit{mdn/content} repository, the documentation of a network explanation was changed based on ChatGPT's response.

\textit{Technical Support} (10 out of 42 samples) - Support queries generally involve seeking assistance with specific technical issues, which may include command prompts generating error messages, application icons not appearing, locating files, performing commit rebase, among others. These do not fall under code generation or review, as they do not involve source code. For example, \textit{PR \#123} from \textit{vrodriguezf/deepvats}, the developer asked ChatGPT how to configure nbdev and git to ignore the outputs of Jupyter Notebooks and only display the code in the repository. 

\textit{Coding Conventions} (5 out of 42 samples) - Developers ask ChatGPT to provide a better understanding of specific terms and patterns employed in a specific programming language. For example, \textit{PR \#22} from \textit{ldez/tagliatelle}, the developer asked ChatGPT about the correct format for field names in TOML files: ``snake\_case", ``camelCase", or ``kebab-case". The developer used ChatGPT's response to support their argument with the reviewer, who had requested a modification of the field names to the 'snake\_case' format."

\textit{Policy} (3 out of 43 samples) -  Developers seek information or advice on managing and defining policies related to various aspects, such as data retention and backup strategies. For example, \textit{PR \#219} from \textit{pwncollege/dojo} repository, the reviewer started a discussion about retention policy for backups using ChatGPT's help, as the developer proposed daily backups to cloud storage.

\subsubsection{Text Review} 

In this category, developers input a text to ChatGPT asking for review and modifications. These texts are often documentation of some system functionality.

\textit{Grammar and Refinement} (10 out of 14 samples) - Developers ask ChatGPT to refine the text, improving grammar or making the text more concise. Generally, these texts are from .md or HTML files within the software, explaining technical aspects or software requirements. For example, \textit{PR \#300} from \textit{daeuniverse/dae} the developer asked ChatGPT to review and refine the grammar of the text, and the output was used to improve the .md file.

\textit{Formatting} (4 out of 14 samples) - This subcategory focuses on reviewing and suggesting improvements in text formatting, including the use of tables, lists, bold, italics, etc., to enhance readability and visual presentation. For example, in \textit{PR \#15455} from the \textit{netdata/netdata} repository, the developer submitted two markdown tables, asking ChatGPT which one presented the content better for a GitHub README. ChatGPT chose one of them and provided justifications for the choice.

\noindent
\begin{center}
\fbox{\begin{minipage}{33em}
\textbf{RQ \#1 Answer: In many cases, developers ask ChatGPT to generate source code, whether by implementing new features or reviewing the developers' own code to help fix bugs and improve overall code quality. However, we also observed instances where developers seek to enhance the quality of text presented to the end-user, solve technical questions, and obtain technical support, particularly concerning errors encountered in the prompt when attempting to execute scripts from the PRs.}

\end{minipage}}
\end{center}
\vspace{.2em}

\subsection{RQ \#2) What is the distribution of the size of the interaction (\#prompts) with ChatGPT to get the desired answer?}

\begin{figure}[ht]
\centering
\includegraphics[width=0.8\textwidth]{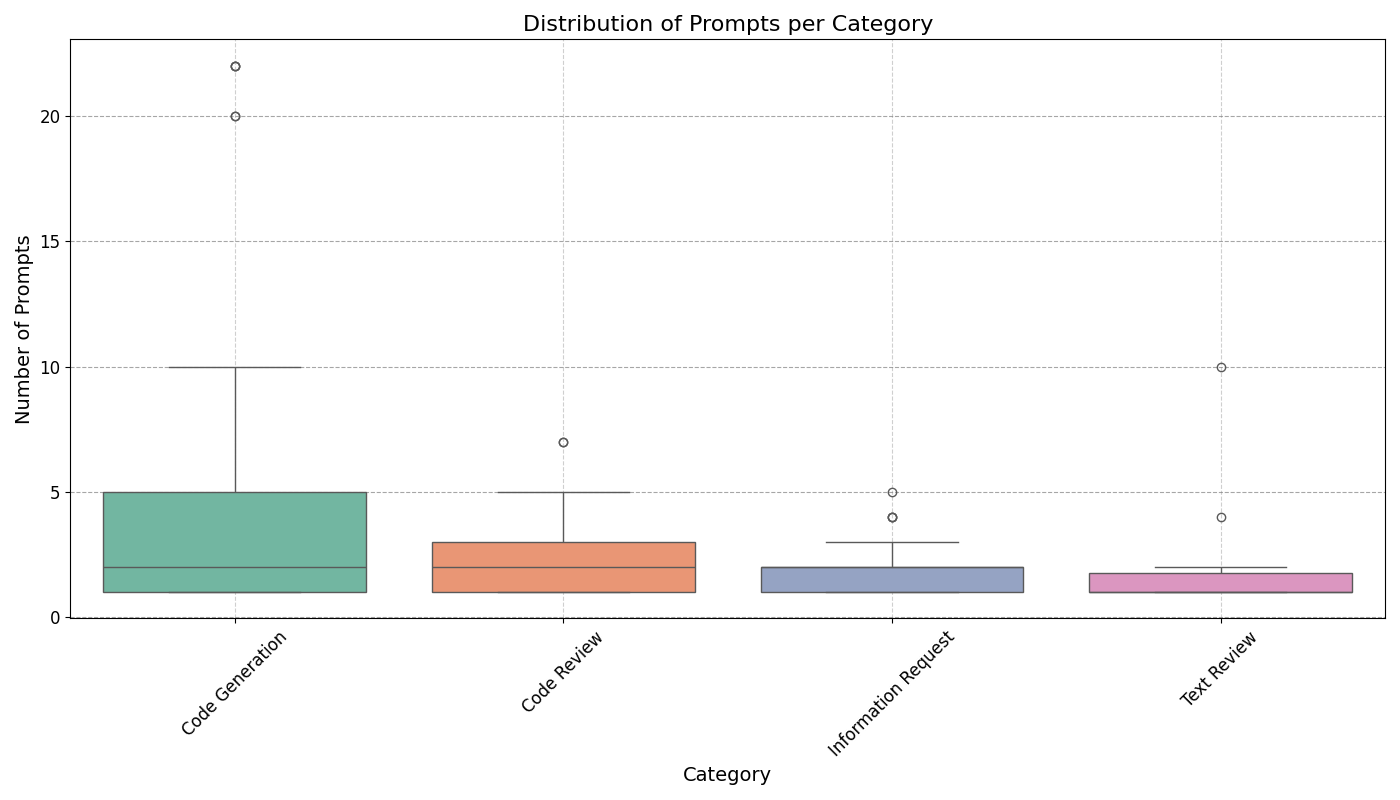}
\vspace{-4mm}
\caption{Distribution of Prompts per Category}
\vspace{-4mm}
\label{fig:prompts}
\end{figure}

Figure \ref{fig:prompts} illustrates the distribution of the number of prompts performed by developers in each context window (i.e., share link) with ChatGPT. The ``Code Generation" and ``Code Review" categories exhibit a wider range of values, suggesting that more prompts were used for each ChatGPT link and with greater variability compared to ``Information Request" and ``Text Review".

%Extracting the \textit{ANOVA} on the number of prompts per ChatGPT link grouped by categories, we obtained \textit{p-value = 0.0003}. This result indicates that there are significant differences between the distributions of prompts per category. The \textit{Tukey Test} found a statistically significant difference between the groups ``Code Generation" and ``Information Request," with a \textit{p-value = 0.0009}, and ``Code Generation" and ``Code Review" \textit{p-value = 0.0062},  indicating that the number of prompts in these categories differs significantly. This result suggests that implementing new features, automating tasks and generate code snippets could result in more effort in prompt engineering for developers to extract the desired answer.

Extracting the \textit{Kruskal-Wallis} on the number of prompts per ChatGPT link grouped by categories, we obtained \textit{p-value = 0.011}. This result indicates that there are significant differences between the distributions of prompts per category. The \textit{Dunn Test} found a statistically significant difference between the groups ``Code Generation" and ``Information Request," with a \textit{p-value = 0.05}, and ``Code Generation" and ``Text Review" \textit{p-value = 0.04},  indicating that the number of prompts in these categories differs significantly. This result suggests that implementing new features, automating tasks and generate code snippets could result in more effort in prompt engineering for developers to extract the desired answer.

\noindent
\begin{center}
\fbox{\begin{minipage}{33em}
\textbf{RQ \#2 Answer: We observed that, in general, developers execute more prompts in the code generation category compared to others. This indicates that generating source code to be merged into a PR may require more extensive prompt engineering work.} 

\end{minipage}}
\end{center}
\vspace{.2em}

\section{Threats to Validity}

In this section, we show some threats to the validity of this study.

\textbf{Generalizability of the Findings}: The findings from RQ \#1 are specifically centered on merged PRs from GitHub, which leverage a relatively lower number of samples due to the recent adoption of ChatGPT in the software development environment. In the near future, with increased maturity in the use of the tool, new samples may indicate new behaviors.

\textbf{Prompt Engineering}: The findings from RQ \#2 could have occurred for reasons beyond the type of request, as the quality of the prompts influences developers' performance during programming tasks.

\section{Related Work}

Three related works have already proposed taxonomies defining the interactions between developers and ChatGPT on GitHub. Tufano et al. \cite{tufano2024} proposed a taxonomy describing the automated tasks requested by developers to ChatGPT, categorizing 467 instances (165 commits, 159 PRs, and 143 issues) that mention the ChatGPT keyword. Our work differs because we restricted our analysis on merged PRs involving at least two individuals (developer and reviewer), and we focuses on what the developer ask in the prompt using the ChatGPT link, instead of classify the types of tasks that mention ChatGPT assistance inside GitHub. 

Couchen et al. \cite{moataz2024} and Hao et al. \cite{hao2024empiricalstudydevelopersshared} both proposed a taxonomy based on shared ChatGPT conversations using the DevGPT dataset \cite{devgpt}, with the former evaluating 243 ChatGPT prompts in PRs and the latter assessing 210 GitHub PRs and 370 issues. In our work, we chose to create our own scripts to mine GitHub PRs instead of using the DevGPT dataset, as we filtered only for merged PRs, repositories with at least 10 stars, involved at least two individuals in a peer review process, and obtained more recent data (up to July 2024 instead of October 2023 as defined in the DevGPT dataset). This approach allowed us to obtain a more refined dataset to achieve our results.

%Nasehi and colleagues \cite{Nasehi2012} studied types of questions in Stack Overflow (SO), and found that the questions would be classified into major five categories: how-to-do, debug-corrective, seeking-something, conceptual and miscellaneous. We could observe that interaction with ChatGPT brought new kind of questions that typically Stack Overflow seems not to be designed for, such as, create documentation or comments, and asking to test or review an piece of code. These new kind of questions shows how developers may use LLMs in a more versatile way.

Although ChatGPT is a recent tool, the evaluation of code generation model quality has attracted  attention from researchers. For instance, recent studies have shown that ChatGPT can successfully generate test cases that induce failures when focused on nuanced aspects \cite{nuances2023}, and it can outperform the state-of-the-art Code Reviewer tool in code refinement tasks \cite{ICSE2024}. While these studies do not propose similar outcomes, they underscore the importance of understanding developer interactions with ChatGPT, as it has proven effective in various tasks. Additionally, Dantas et al. found that ChatGPT can generate didactic and high-quality code snippets that are often easier to read than those produced by humans on Stack Overflow \cite{dantas2023}.

\section{Conclusion}

In this study, we aim to qualitatively understand what developers request from ChatGPT in the context of PRs that contain modifications accepted by reviewers. We observed that more than half of the requests involve generating source code as output, whether it be implementing a new feature requested by the developer, improving existing code, fixing bugs, explaining what the code does, or even evaluating code´s performance. However, there is also a considerable number of samples where developers ask ChatGPT to enhance the quality of the text presented in the front-end application, as well as provide support for commands or explain complex software engineering concepts. In most cases, we found that the solutions offered by ChatGPT contributed to the PR, either in the committed source code or as clarifications within the PR discussions between developers and reviewers. We also noted that when developers request code generation, the number of prompts increases significantly, indicating that producing useful code for the production environment may require additional prompt engineering.

This work has several areas for future improvements. For example, instead of solely examining what developers ask ChatGPT, one could also evaluate the reasons behind the questions, i.e., the context in which they were requested within merged PRs. Another interesting aspect would be to analyze rejected PRs that included assistance from ChatGPT, aiming to classify scenarios where ChatGPT's help may have been insufficient. Additionally, prompt outliers, such as instances where developers needed to ask multiple questions to obtain their desired answers, should also be analyzed.

\bibliographystyle{sbc}
\bibliography{sbc-template}

\end{document}